\RequirePackage{fix-cm}
\documentclass[twocolumn]{svjour3}
\smartqed  
\usepackage{graphicx}
\begin{document}

\title{Pressure effects and orbital characters in 
cuprate and carbon-based superconductors
}

\titlerunning{Pressure effects and orbital characters in 
cuprate and carbon superconductors}        

\author{Hideo~Aoki \and Toshikaze~Kariyado}


\institute{H.~Aoki \at
Department of Physics,\\
University of Tokyo,\\
Hongo, Tokyo 113-0033, Japan \\
\email{aoki@phys.s.u-tokyo.ac.jp}           
\and
T.~Kariyado \at
Division of Physics, University of Tsukuba,\\
Tsukuba, Ibaraki 305-8571, Japan \\
\email{kariyado.t.gf@u.tsukuba.ac.jp}
}

\date{Received: date / Accepted: date}

\maketitle

\begin{abstract}
Pressure effect is overviewed for 
the cuprates and carbon-based superconductors, with an emphasis 
on how their orbital characters are modified by pressure.  
For the  high-Tc cuprates, we start from an observation for ambient
 pressure that, on top of the main  
orbital ($dx^2-y^2$), a hybridization with the second ($dz^2$) orbital
 around the Fermi energy significantly affects $T_c$ in the
 spin-fluctuation mediated pairing, where the hybridization is dominated
 by material parameters. We can then show that applying pressures along
 a, b axes enhances 
Tc while a c axis pressure suppresses Tc, where not only the $dz^2$ 
hybridization but also Cu($4s$) hybridization exert an effect.  
For the multi-layer cuprates, inter-layer pair hopping is suggested to be 
important, which may contribute to pressure effect. 
Pressure effect is also interesting in a recently discovered aromatic
 family of superconductors (picene, etc).  There, we have again
 multi-band systems, which in this case derive from different molecular
 orbitals.  The Fermi surface is 
an intriguing composite of different pockets/sheets having different
 dimensionalities arising from anisotropic transfers between the
 molecular orbitals, and pressure effects should be an important probe
 of these. 
\keywords{Cuprate \and carbon-based superconductors \and pressure effect}
\end{abstract}

\section{Introduction}
Pressure effect is one of the most interesting and direct probes and 
ways to control of the relation between electronic structure and
superconductivity.  One starting point 
can be an observation that, in the  high-Tc cuprates at ambient pressure, 
there exists significant material dependence of $T_c$, which remained an
important puzzle even within the single-layer family.  A recent work by
the present author and
coworkers\cite{aoki,sakakibara,sakakibara_pressure} has demonstrated,
with a two-orbital model, that, while the usual wisdom is to consider
the cuprate as a one-band ($dx^2-y^2$) system, a hybridization with the
second ($dz^2$) band around the Fermi energy significantly affects $T_c$
in the spin-fluctuation mediated pairing.  There, the larger the energy
offset ($\Delta E$) between the two orbitals (i.e., smaller the band
mixing), the higher the $T_c$.  

We can then follow this line of approach to show that applying pressures can 
significantly affect Tc, through the pressure dependence of the 
$dz^2$ hybridization.  
If we turn to the multi-layer cuprates, inter-layer pair hopping is
suggested to be important\cite{nishiguchi}, which may have an
interesting effect on pressure dependence.

Pressure effect is also interesting for carbon-based superconductors, 
typically fullerene systems, where both of the electron-electron and 
electron-phonon interactions are strong.  We can then discuss the 
Tc dome against the interactions.\cite{murakami} 
In a recently discovered aromatic family of superconductors (picene,
etc)\cite{Mitsuhashi2010}, we have again multi-band systems\cite{aoki},
which in these class of materials derive from multiple molecular
orbitals.  As a result the Fermi surface is an intriguing composite of
different pockets/sheets having different dimensionalities arising from
anisotropic transfers between the molecular orbitals, and pressure
effects should be an important probe of these. 

\section{Cuprates}

\begin{figure*}[htbp]
 \begin{center}
  \includegraphics[width=13cm]{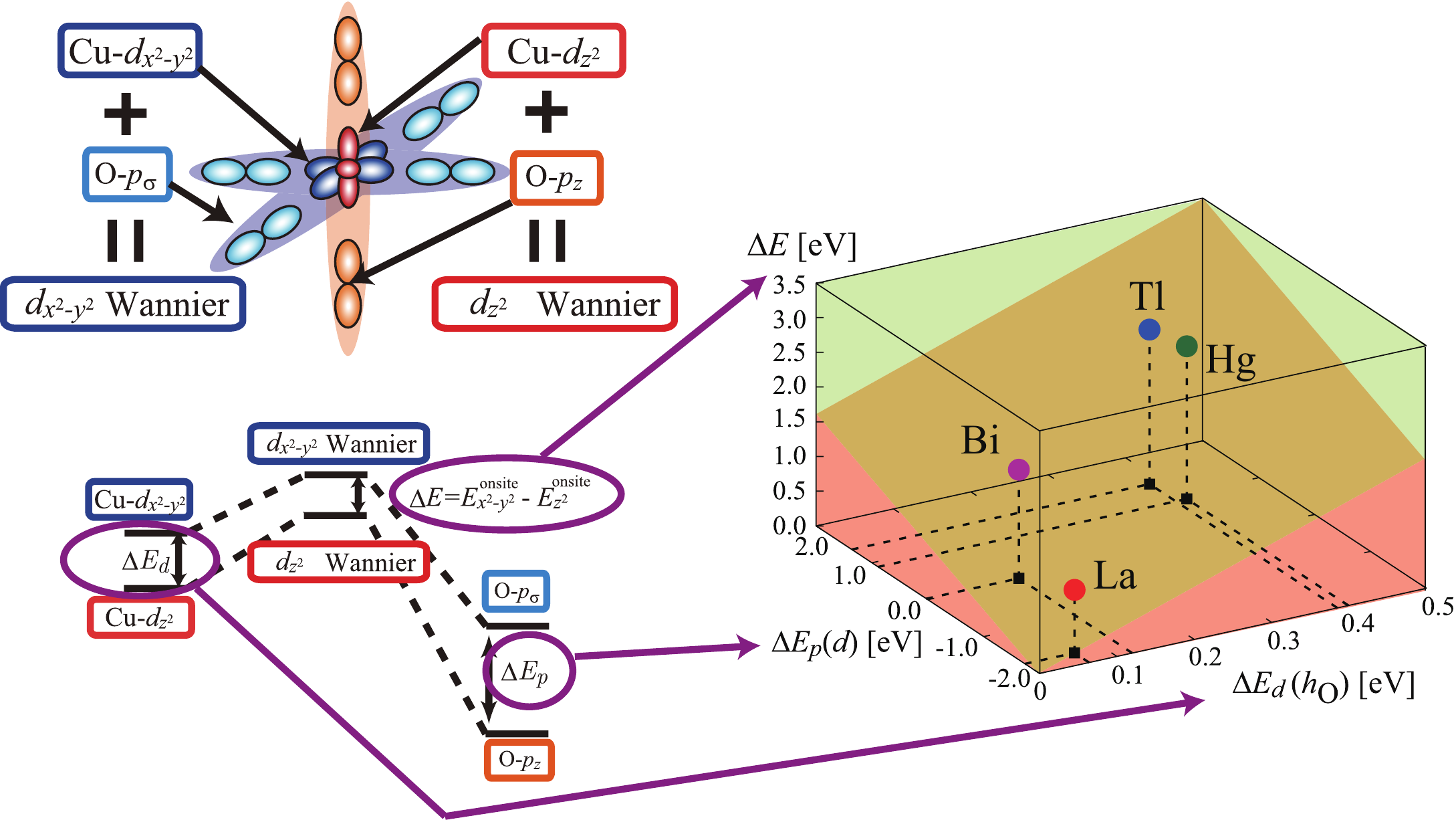}
  \caption{
Left top: The main components for each of the two 
Wannier orbitals in the cuprate. 
Left bottom: Level offsets $\Delta E$ of the two Wannier 
orbitals as derived from 
the d-level offset $\Delta E_d$ and the p-level offset $\Delta E_p$.  
Right: 
$\Delta E$ plotted against $\Delta E_d$ and $\Delta E_p$ 
for various single-layered cuprates.  
An oblique plane indicates a 
rough correlation between $\Delta E$ and 
($\Delta E_p, \Delta E_d$). After \cite{sakakibara}.  
}
  \label{levels}
 \end{center}
\end{figure*}
In the physics of high-$T_c$ cuprates, optimizing their $T_c$ 
remains a fundamental yet still open problem.   Empirically, important 
parameters that control $T_c$ have been identified to be 
chemical composition, structural parameters, the number of layers, etc, 
besides the doping concentration.  
Specifically, 
several key structural parameters have been suggested: 
the bond length 
between copper and in-plane oxygen\cite{Jorgensen,Bianconi}, and 
the Cu-apical oxygen distance ($h_{\rm O}$)
\cite{Maekawa,Andersen,Feiner,Pavarini,Kotliar,Weber,Takimoto,sakakibara}. 

As mentioned in Introduction, 
Refs.\cite{sakakibara} conclude that 
larger the level offset $\Delta E$ between the $dx^2-y^2$ and $dz^2$ 
Wannier orbitals, higher the $T_c$, where $\Delta E$ is governed by the
apical-oxygen height and the inter-layer distance.  
The key material parameters in the cuprates 
are: 
the apical oxygen height ($h_{\rm{O}}$) above the CuO$_2$ plane and the
separation ($d$) between the CuO$_2$ planes determine the level offset
$\Delta E$ between the Cu($dx^2-y^2$) and Cu($dz^2$) {\it Wannier}
orbitals through $\Delta E_p$ (offset between O($p_z$) and
O($p_{\sigma}$) and $\Delta E_d$ (offset between atomic Cu($dx^2-y^2$)
and Cu($dz^2$)).  This captures the material dependence of $T_c$,
covering a broad range of single-layer cuprates (La, Hg, Bi, and Tl).
These are depicted in Fig.~\ref{levels}.

Pressure effect then is an ideal 
{\it in situ} way to probe the structure-dependence of $T_c$. 
So far we know for the cuprates that: 
(i) $T_c$ tends to be enhanced under hydrostatic pressure for pressure $< 30$ GPa \cite{Klehe,Gao}, while 
(ii) uniaxial pressures produce anisotropic responses of $T_c$, namely, 
an $a$-axis 
compression generally raises $T_c$ 
while $c$-axis compression has an opposite 
effect
\cite{Hardy,Gugenberger,Meingast}.  Moreover, 
the magnitude of the pressure coefficient 
tends to be smaller for materials 
having higher $T_c$'s\cite{Hardy}.

\begin{figure}[htbp]
 \begin{center}
  \includegraphics[width=8cm]{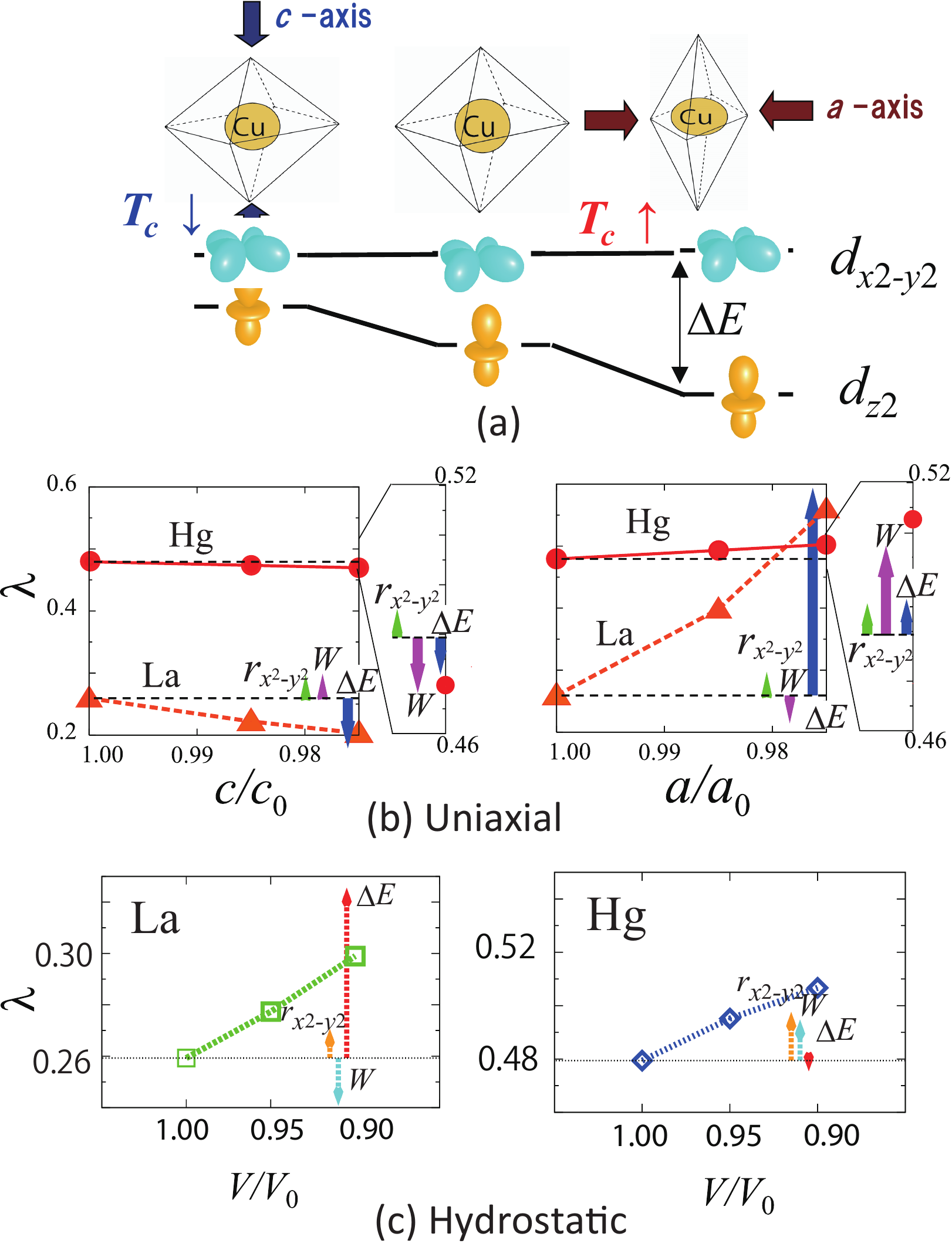}
  \caption{(a) A schematic effect of uniaxial compression 
along c or a axis.  
(b) For uniaxial compressions for La(Hg) cuprates, 
the eigenvalue $\lambda$ of the Eliashberg equation is plotted against $c/c_0$ (left panel) or  $a/a_0$ (right). 
Triangles (circles) indicate the result for the La (Hg) cuprates.
Arrows depict the contributions to the pressure effect 
from $\Delta E$, $W$, and $r_{x^2-y^2}\equiv (|t_2|+|t_3|)/|t_1|$, respectively.  
(c) For hydrostatic compression, $\lambda$ is plotted against the volume 
$V/V_0$.   
Arrows indicate various contributions as in (b). After \cite{sakakibara_pressure}.  
}
  \label{pressure_fig2}
 \end{center}
\end{figure}

Basically, the effects of uniaxial 
pressures come from the pressure-dependence 
of $\Delta E$ affected by the crystal field, 
as schematically shown in 
Fig.~\ref{pressure_fig2}(a), where a-axis compression, 
which enhances the apex O height, raises Tc.   
More precisely, however, 
while the variation of $T_c$ under pressure 
is indeed affected by $\Delta E$, especially in the relatively low-$T_c$ 
cuprates, the higher-$T_c$ cuprates such as 
HgBa$_2$CuO$_4$ (Hg1201) with large $\Delta E$ has a 
smaller $T_c$ variation against pressure.  
We can also show that 
one factor contributing the pressure effect 
is, on top of $dx^2-y^2$ and $dz^2$ orbitals, 
Cu($4s$) orbital can be relevant, whose level 
is raised with pressure, 
resulting in a better nested Fermi surface. 
This is shown in a model comprising 
all of the $dx^2-y^2, dz^2$ and $4s$ orbitals explicitly.  
This, along with the increase in the band width, is shown to 
cause a higher $T_c$ under pressure.   

So we display the theoretical result, for La and Hg cuprates, 
for the dependence of the eigenvalue of the 
Eliashberg equation (a measure of Tc) 
on uniaxial pressures in Fig.~\ref{pressure_fig2}(b) 
or on the hydrostatic pressure in Fig.~\ref{pressure_fig2}(c).  
There we have decomposed the pressure effects 
into three 
contributions from the change of $\Delta E$ due to pressure, 
change of $W$, and change of the ratio, $r_{x^2-y^2} \equiv 
(|t_2|+|t_3|)/|t_1|$, of the further neighbour 
hoppings to the nearest neighbour one that determines the 
warping of the Fermi surface. 
For details see Ref.\cite{sakakibara_pressure}.

\begin{figure}[htbp]
 \begin{center}
  \includegraphics[width=8cm]{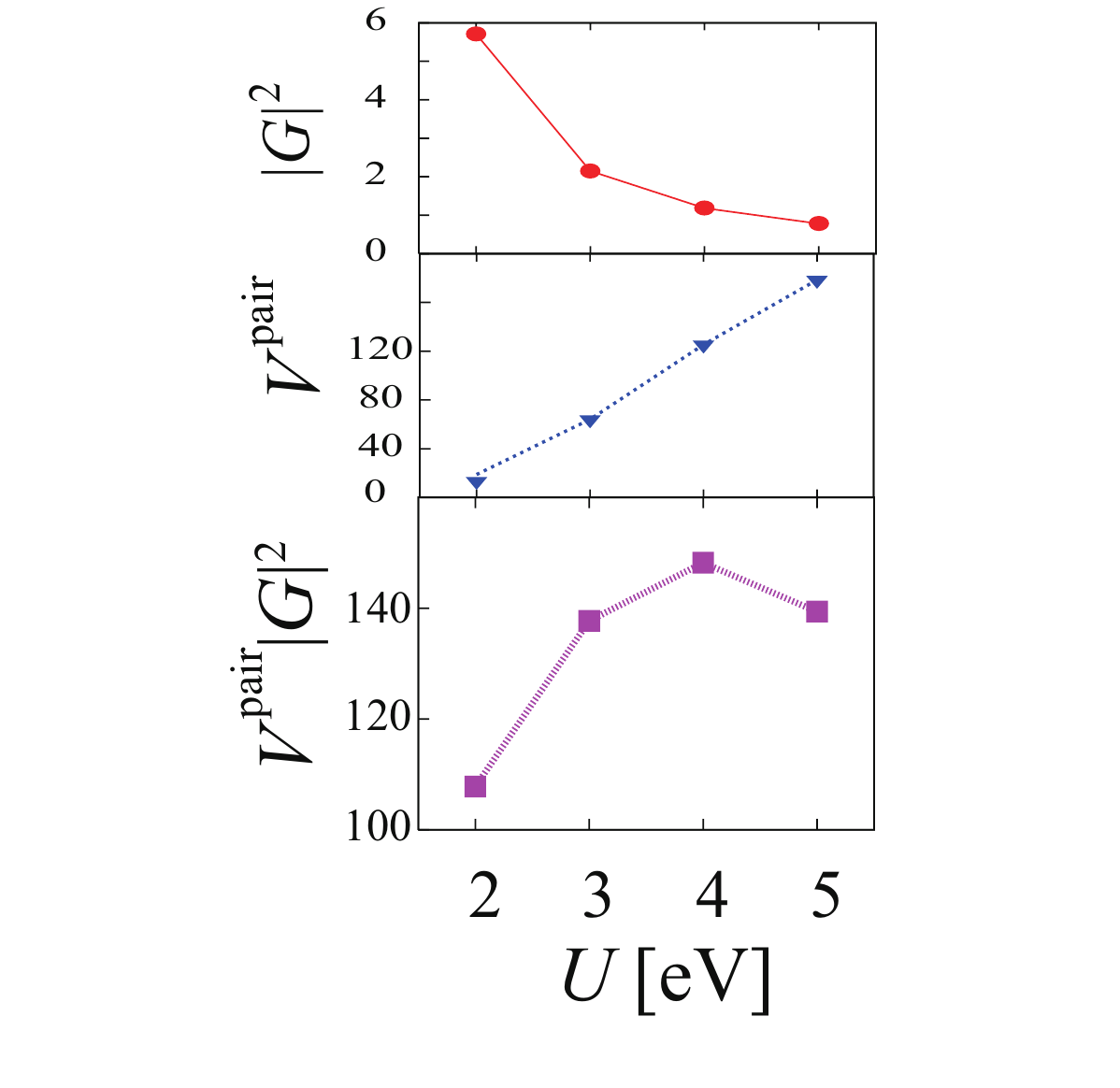}
  \caption{
$U$ dependence of the absolute value of the renormalized
 Green's function squared $|G|^2$ at $(\vec{k},i\omega)=(\pi,0,i\pi k_BT)$,
the effective pairing interaction 
$V^{\rm pair}$ at $(\vec{q},i\omega)=(\pi,\pi,0)$ 
and the product $V^{\rm pair}|G|^2$.\cite{sakakibara_pressure}. }
  \label{vgg}
 \end{center}
\end{figure}

As for the contribution from $W$ a hydrostatic pressure 
increases $W$.  However, in Fig.~\ref{pressure_fig2}(c) we can notice 
that the increase in $W$ 
results in an enhancement of $\lambda$ in Hg, while the 
opposite occurs for La.  
If we vary $U$ with a fixed band width $W$,  
the absolute value of the renormalized
 Green's function squared $|G|^2$ 
at $(\vec{k},i\omega)=(\pi,0,i\pi k_BT)$ 
monotonically decreases 
with $U$ 
as in Fig.~\ref{vgg} (for the Hg compound 
in this figure) due to an increased self-energy.  
On the other hand, 
the pairing interaction $V^{\rm pair}$ increases with $U$ 
because the spin fluctuations develop monotonically.
Consequently, $V^{\rm pair}|G|^2$, 
a rough measure of the eigenvalue of the Eliashberg equation 
for $d-$wave superconductivity, exhibits a peak. 
If we repeat the calculation for the La cuprate, 
we can show that the Hg cuprate is located on the right of 
the peak, while the La cuprate on the left.  This explains 
the opposite contributions for the $W$ effect.

\begin{figure}[htbp]
 \begin{center}
  \includegraphics[width=8cm]{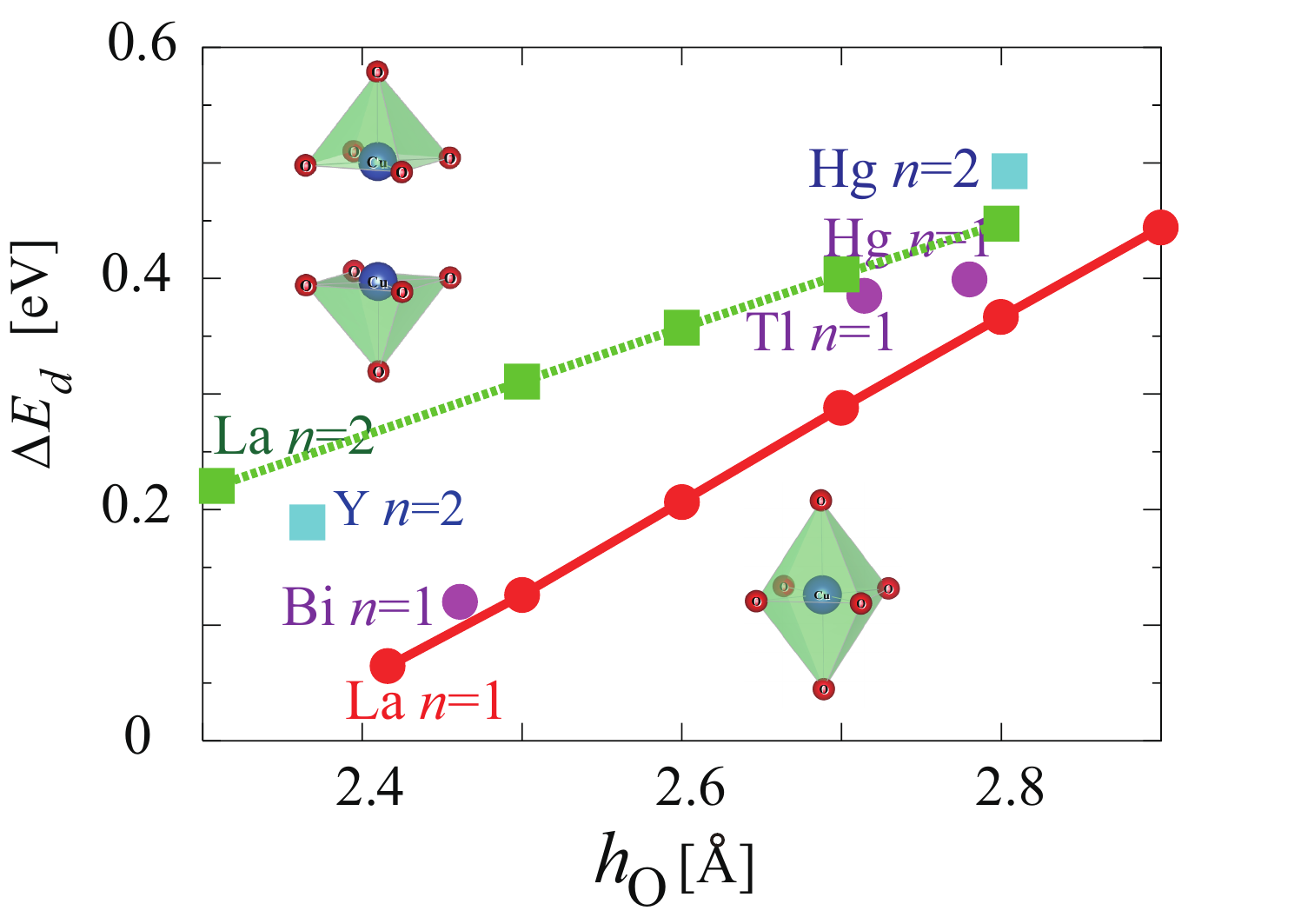}
  \caption{
Relationship between the apical oxygen height $h_{\rm O}$[\AA] and
the level difference $\Delta E_d$ [eV]. The squares(circles) 
represent bi-layer pyramidal (single-layer octahedral) system 
with the materials indicated.\cite{sakakibara_pressure}.   
The dashed(solid) line is the result when $h_{\rm O}$ is varied hypothetically.}
  \label{ho-ed}
 \end{center}
\end{figure}

Since the important units in the cuprates come in two flavours, 
CuO octahedron and CuO pyramid, the difference in the effect 
of the apex oxygen height between them is intriguing.  
Figure~\ref{ho-ed} shows the relationship between 
the apical oxygen height $h_{\rm O}$ and $\Delta E_d$ 
(the energy difference between the 
$dx^2-y^2$ and the $dz^2$ atomic orbitals).
We can see that, while 
$\Delta E_d$ is positively correlated with the apical oxygen height
in both octahedral and pyramidal systems.  More precisely, 
while the value of $\Delta E_d$ is significantly greater for 
pyramids, dependence on $h_{\rm O}$ is smaller 
for pyramids with only one apical O per unit. 
This may be one reason why $T_c$ increases in spite of 
somewhat larger reduction in $h_{\rm O}$ 
in bi-layer system than in single-layer system 
under hydrostatic pressure\cite{Jorgensen}.

The orbital effects can be unified into a picture in which higher $T_c$ can be 
achieved by the ``distillation''\cite{sakakibara_pressure} of the main (i.e., 
$dx^2-y^2$) band, 
namely, the higher the one-band character the 
better. This may also possibly help materials-design for higher Tc.

\begin{figure}[htbp]
 \begin{center}
  \includegraphics[width=8cm]{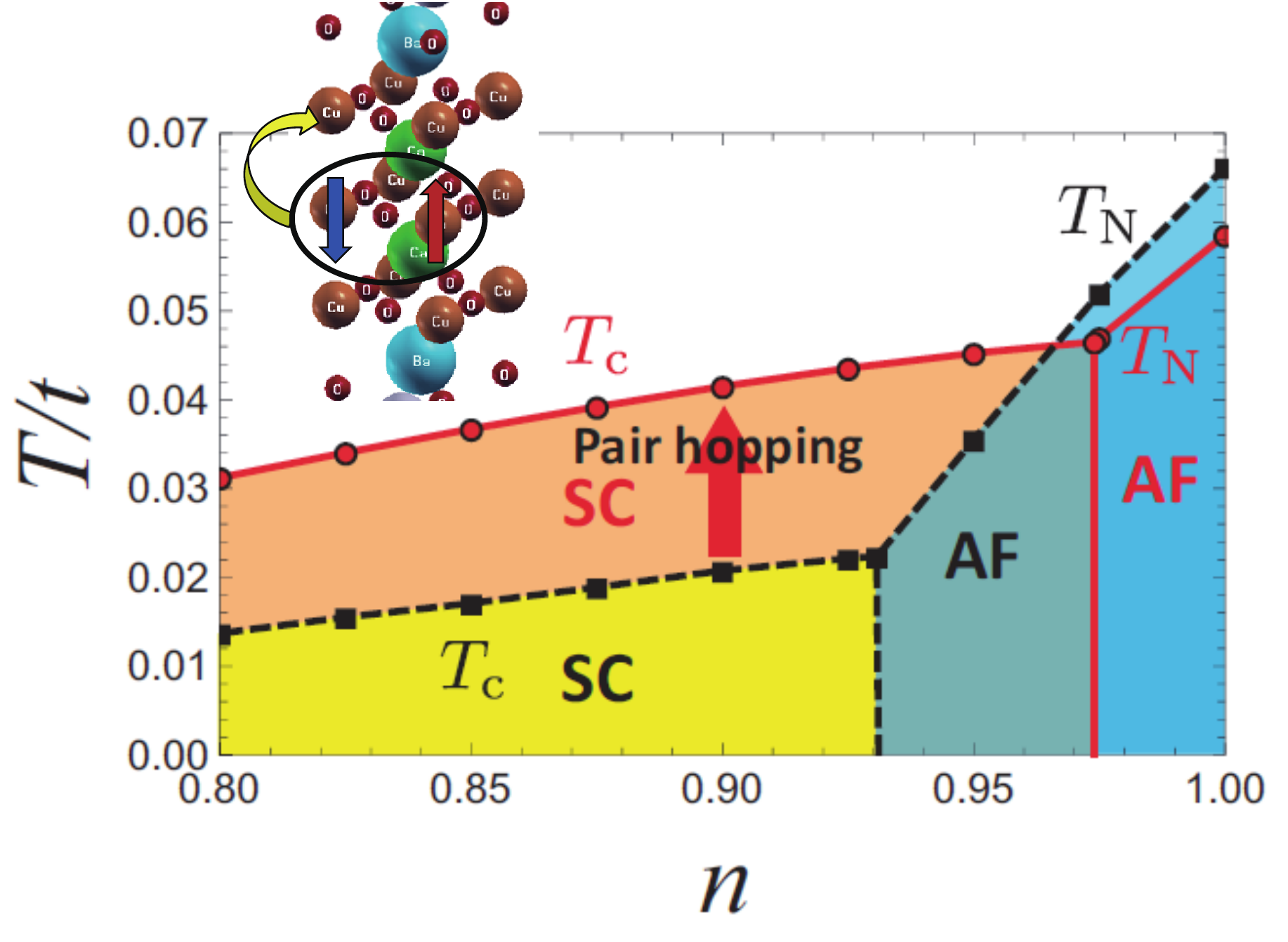}
  \caption{
Phase diagram on $T$ and $n$ (carrier concentration) 
for the double-layer system with (red  lines) and without (black) interlayer pair hopping processes. 
$T_{\rm c}$ is SC transition temperature while $T_{\rm N}$ AF transition (Ne\'{e}l) temperature. Inset schematically depicts the 
interlayer hopping of off-site pairs.\cite{nishiguchi} 
}
  \label{pairhop}
 \end{center}
\end{figure}

In the cuprate family, the record holder of the highest $T_{\rm c}$ 
is still the multilayered cuprates that have $n$ CuO$_{2}$ planes in a unit cell, 
typically the Hg-based HgBa$_2$ 
Ca$_{n-1}$Cu$_n$O$_{2n+2+\delta}$ (Hg-$12(n-1)n$), 
where $T_{\rm c}$ increases for $n=1-3$ and decreases for $n \geq 4$.\cite{Schilling93} 
While there have been several theoretical studies for multilayered cuprates, 
Nishiguchi et al have recently examined 
various microscopic processes.\cite{nishiguchi} 
The interlayer one-electron hopping has little effects on the band structure, 
and they have focused on the interlayer pair hopping. 
The superconductivity in a double-layer Hubbard model with and without
the interlayer pair hopping, 
as studied by solving the Eliashberg equation with the fluctuation
exchange approximation, 
reveals that the interlayer pair hopping (Fig.~\ref{pairhop}, inset) acts
to increase the pairing interaction and the self-energy simultaneously, 
but that the former effect supersedes the latter and enhances the
superconductivity (Fig.~\ref{pairhop}). 
Study of the triple-layer case with the interlayer pair hopping reveals that 
the superconductivity is further enhanced but tends to be saturated
toward the triple-layer case.  

\section{Fullerene superconductor}

There is a class of superconductors in which 
the electron-electron Coulomb repulsion and the electron-phonon coupling are 
strong at the same time.  This most typically applies to 
alkali-doped fullerides.  Although the pairing is an $s$-wave, 
Tc has recently been found to be dome-shaped and 
sits next to an antiferromagnetic phase in the temperature-pressure phase diagram\cite{c60_1,c60_2,c60_3}.  Thus an interplay of 
the electron-electron and electron-phonon interactions becomes an interesting 
problem.  This may also apply to 
the aromatic superconductors in the next section.  

We can tackle this problem with the Holstein-Hubbard model, which is a
simple model that incorporates both electron-electron and
electron-phonon interactions. The model is characterized by the energy
$\omega_0$ of dispersionless (Einstein) phonons, the on-site Hubbard
interaction $U$ and the electron-phonon coupling $g$.  
There is a body of works which investigates the competition between the
two interactions in this model. The study of the one-dimensional case
based on the density matrix renormalization group (DMRG) technique or
quantum Monte Carlo analysis has revealed some general
features.\cite{1dimdmrg1,1dimdmrg3,sse1} 
In the opposite limit of infinite spatial dimensions, $D=\infty$, where
the dynamical mean-field theory (DMFT) becomes exact,
the effect of the competition between the two kinds of interactions have
been
studied~\cite{HHdmft4,HHnrg,HHdmft5,HHdmft6,HHdmft1,HHdmft2,HHdmft3,HHdmft7}. Two-dimensional 
system is also studied.\cite{nowadnick}

However, an important question remains as 
follows: a usual wisdom in capturing the system is to regard it as having an 
effective interaction $U_{\rm eff} \equiv U-\lambda$, where $\lambda$ is
the static effective electron-electron interaction mediated by phonons.  
However, this is only valid 
in the antiadiabatic limit for 
$\omega_0\rightarrow\infty$ where the phonon-mediated 
interaction becomes 
non-retarded, so that the real question is to what extent this
approximation remains valid when we vary 
$U$ and/or $\omega_0$.   In other words how can we go beyond 
Migdal theorem\cite{sc3,sc4} or the MacMillan equation.\cite{sc5}
In the fullerene superconductor, $U$, $\lambda$, and $\omega_0$ are all
comparable to the electronic bandwidth $W$, so the question becomes
real.  

\begin{figure}[htbp]
 \begin{center}
  \includegraphics[width=8cm]{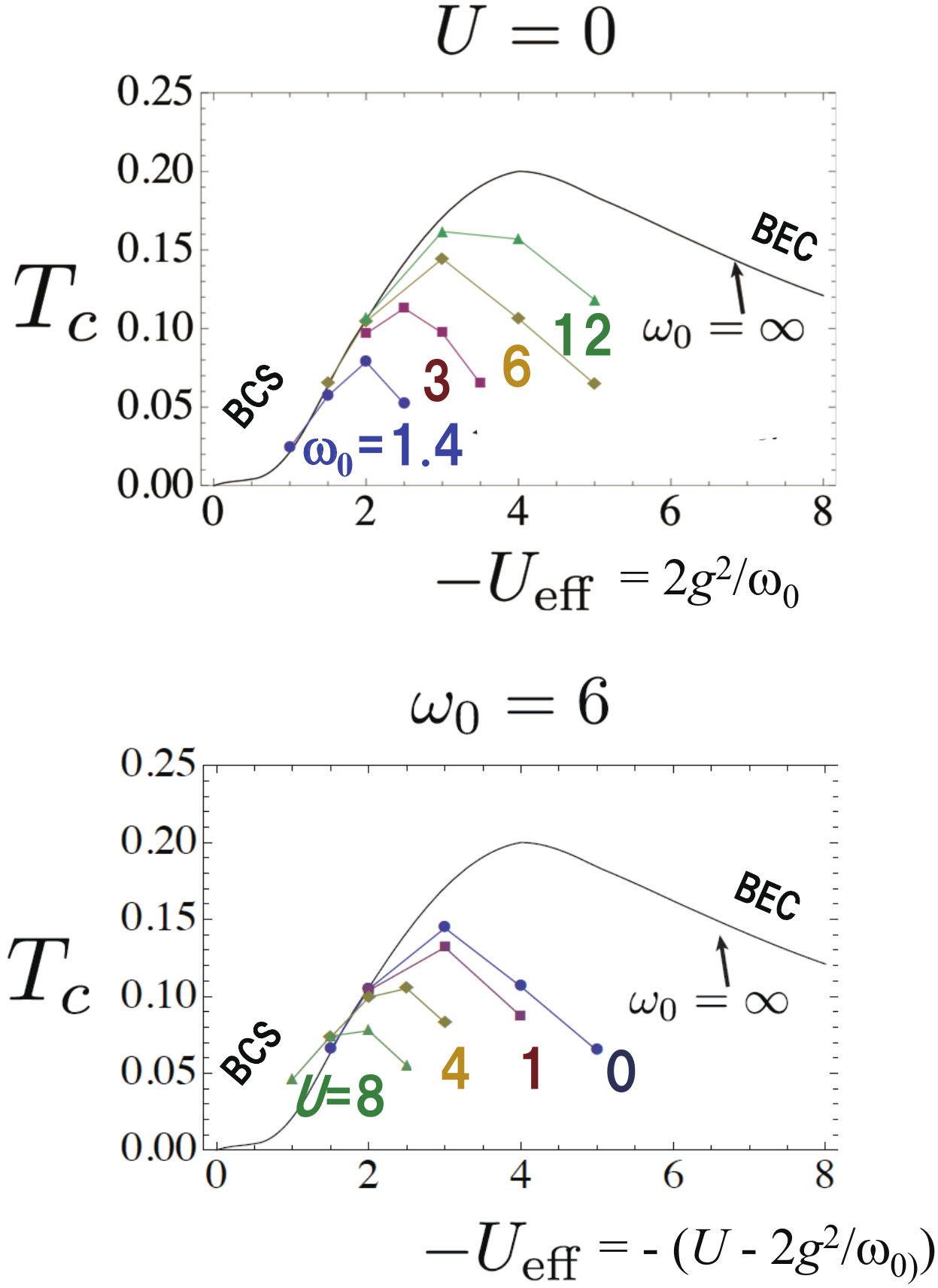}
  \caption{
$T_C$ against $-U_{\mathrm{eff}}$ for various values of the phonon 
energy $\omega_0$ with the 
Hubbard $U=0$ 
(top panel), and for various values of $U$ with $\omega_0=6.0$ (bottom), with 
the hopping integral $t$ as the unit of energy.  
 The lines connecting symbols are guides for the eye, while 
the black curve in each panel indicates $T_c$ in the attractive Hubbard model. 
After \cite{murakami}.  
}
  \label{HH}
 \end{center}
\end{figure}

We have studied the  Holstein-Hubbard model, at half-filling, 
with the dynamical mean-field theory (DMFT), with a continuous-time
quantum Monte Carlo impurity solver, which enables us to work 
in the regime where $U$, $\lambda$, and $\omega_0$ are comparable to the
bandwidth.\cite{murakami} 
We show in Fig.~\ref{HH} that the phonon-induced retardation 
or the Coulomb repulsion have the effect of significantly decreasing and 
shifting the $T_c$ dome against $U_{\rm eff}$. 
In order to understand and interpret the observed behavior we can
introduce an effective static model derived from a Lang-Firsov
transformation.
As for the pressure effects, the quantities $\omega_0$, $U$ and $\lambda$ 
should be affected by pressure, which is an interesting future problem to 
elaborate.  In real fullerides there is also a complication coming from 
the Jahn-Teller degree of freedom.\cite{c60_2}

\section{Aromatic superconductor}

Let us move on to the pressure effects on the aromatic hydrocarbon
superconductors, especially picene
superconductors\cite{Mitsuhashi2010}. Potassium (K)
doped picene has been attracting much attention after the 
experimental report of 
superconductivity with transition temperature Tc=18 K, which 
is very high as an organic 
material. Furthermore, there is a very recent report for a striking pressure
effect, where Tc is raised to about 30 K by applying
pressure $\sim 1$ GPa\cite{Kambe2012}. At present, neither the pressure effect 
nor is even the
superconducting mechanism itself well understood
theoretically. If the picene superconductor obeys conventional BCS
theory, the density of states (DOS) at the Fermi energy will be an
important factor determining Tc. The positive pressure dependence of Tc
seems to contradict with a conventional mechanism, since applying
a pressure is likely to decrease DOS. Specifically, we have to 
carefully examine how the pressure affects the electronic 
structure, since the theoretical band structure of
picene superconductors, being multi-orbital systems, 
is rather complex\cite{Kosugi2011}. This is important, since, 
if an unconventional
mechanism, such
as the electron-electron correlation, is at work, Tc will be sensitive to 
the 
Fermi surface structure.  Thus, in the following, we show
some results on the electronic structure under pressure.

Here electronic and crystalline structures of K-doped picene
crystals are investigated based on the density functional theory (DFT)
with local density approximation (LDA) adopted for
exchange-correlation functionals. The actual calculations are
performed using Quantum Espresso package\cite{QuantumEspresso}. We
concentrate on
the case in which three K atoms are doped per a picene molecule, 
for which the superconductivity is reported. 
At ambient pressure, the
lattice constants are fixed to the most updated experimental
values\cite{Kambe2012} in the present study, and the internal coordinates are relaxed by minimizing
calculated forces on each atom. 
For the pressure effect, the 
calculation is done for an ``artificial''
pressure introduced as decreased lattice parameters, as a first step towards 
understanding of pressure effects in picene superconductors.
Thus the pressure effect is simulated by 
repeating the calculation with reduced lattice
parameters. For simplicity, lattice constants in three directions are
scaled by a common scaling factor, which means that the anisotropy in the 
bulk modulus is neglected. We carried out calculations with scaling
factors 100\% (experimental lattice constants at ambient pressure), 
down to 95\%.   
Ideally, a fully theoretical method is desirable, where 
the lattice constants are relaxed within a theoretical 
structural optimization under pressure. However, precise prediction
of the lattice constants within DFT is difficult for the present target
material.  
There are several reasons for this.  A theoretical factor is that 
we should have, especially for the undoped molecular crystal, 
a calculation with the van der Waals interaction taken into account, 
but the method is
still under improvements. Once the lattice
constants are given, the difference between LDA and generalized-gradient
approximation (GGA) is less significant for internal coordinates. 
A second, experimental factor here is 
that there is no definite experimental lattice constants, 
which is mainly because the superconducting fraction is 
very tiny ($\sim 1$ per cent), even at ambient pressure.  
These are why we have opted for the artificial treatment
of pressure effects, and leave more
sophisticated works as future problems.

Even when the lattice constants are fixed and only internal coordinates 
are relaxed, calculation is still hard, since the organic system 
has many metastable structures. For example, there are several possible ways 
for putting the doped 
K atoms (three per molecule) 
into the molecular crystal. Here we focus on, following the
previous theoretical work by Kosugi et al\cite{Kosugi2011}, 
two kinds of K atom arrangements, which we call 
K$_{3}$Picene and K$_2$K$_1$Picene structures, 
as most probable candidates. In K$_3$Picene, all the K atoms are doped
within the layer of picene molecules arranged in a herringbone
pattern. On the other hand, K$_2$K$_1$Picene has one K
atom resides in the interlayer region while the other two 
in the intralayer region.  In
practice, we start internal coordinate relaxation from the 
K$_3$- or K$_2$K$_1$-structure for each set of lattice
constants.

\begin{figure}[htbp]
 \begin{center}
  \includegraphics[width=8cm]{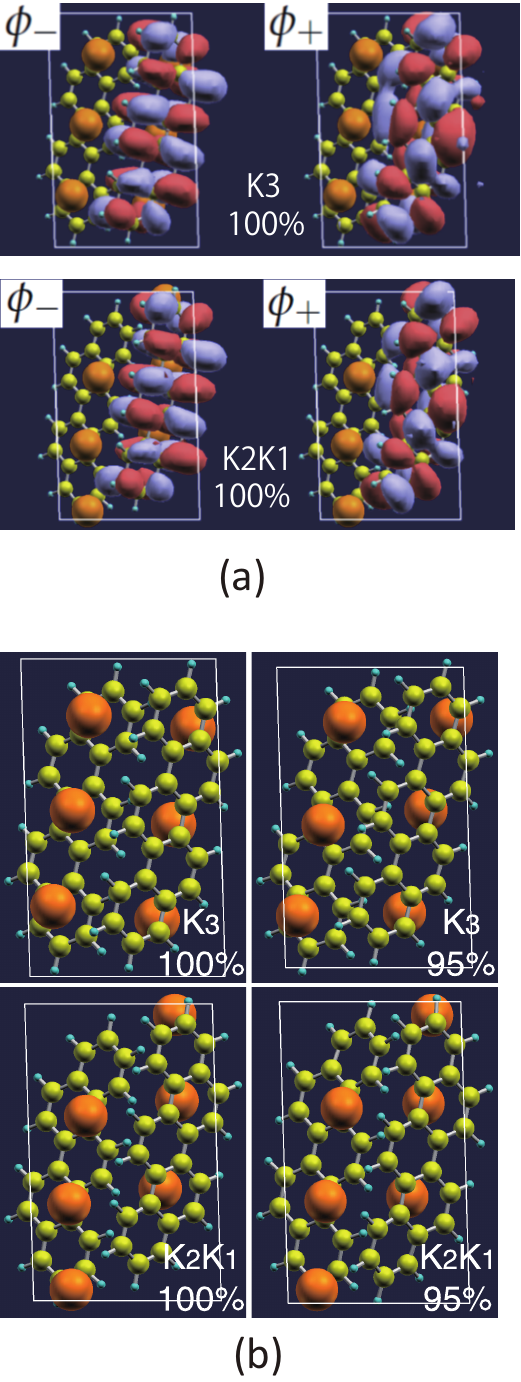}
  \caption{(a) Two Wannier orbitals relevant to the conduction band 
for K$_{3}$Picene and K$_2$K$_1$Picene at ambient pressure.  
(b) Theoretically optimized crystal structures at ambient pressure 
(100\%) and for the 95\% reduced lattice constants.  
Each panel displaying a unit cell is a side view of the molecular layer.}
  \label{picene_fig1}
 \end{center}
\end{figure}

Fig.~\ref{picene_fig1}(a) shows the calculated crystal structures 
along with the two Wannier functions (basically the LUMO and LUMO+1 of 
the picene molecule) for the conduction 
band (which comprises four bands) at ambient pressure.  
The ambient 
pressure result, with the experimentally updated 
lattice constants, shows that K$_3$- and K$_2$K$_1$ continue 
to be the (meta)stable structures.  If we turn to 
Fig.~\ref{picene_fig1}(b), which displays the change of the structure 
when the lattice constants are reduced to 95\%, 
we can see that the main features of K$_3$- and K$_2$K$_1$-structures are 
respectively retained even with the shrunk lattice.   
However, a closer examination 
reveals that there are significant differences in 
the dopant positions and the relative angle between two picene molecules.

\begin{figure}[htbp]
 \begin{center}
  \includegraphics[width=8cm]{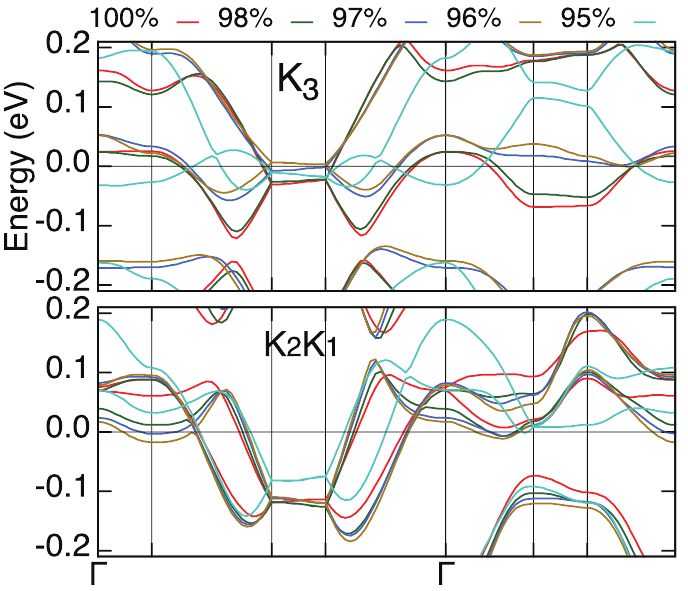}
  \caption{Change in the band structure 
as the lattice constants are reduced down to 95\% 
for K$_{3}$Picene and K$_2$K$_1$Picene. 
Energies are measured from the Fermi energy.}
  \label{picene_fig2}
 \end{center}
\end{figure}
The calculated band structures are plotted in Fig.~\ref{picene_fig2}. 
We first note that
the obtained band structures are different from the previous theoretical
work\cite{Kosugi2011} even at ambient pressure.  This 
naturally comes from the adaptation in the present
calculation of the updated experimental lattice
constants\cite{Kambe2012}. Basically 
the c-axis is elongated in the update, so that the
band structure becomes more two-dimensional and simpler compared with
the previous calculation. If we turn to the pressure effect, 
a notable feature in Fig.~\ref{picene_fig2} is that the band structure
changes in a 
nonmonotonic manner against pressure, especially for 
K$_3$-structure. This comes from the dopant positions and relative
angle between molecules that change nonmonotonically against pressure, 
where the band
structure is sensitive to the change.  
Although a discussion of 
Tc would require a more sophisticated treatment of the 
pressure, a message here is that a decrease in the lattice constants 
can exert an effect beyond a simple expansion of band width.

\begin{figure}[htbp]
 \begin{center}
  \includegraphics[width=8cm]{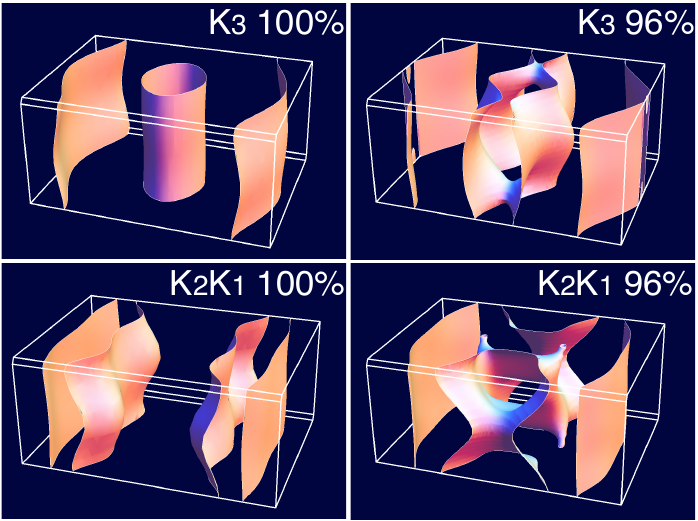}
  \caption{Change in the Fermi surface as the 
lattice constants are reduced down to 96\% 
for K$_{3}$Picene and K$_2$K$_1$Picene. }
  \label{picene_fig3}
 \end{center}
\end{figure}

This fact is highlighted in the Fermi surfaces in Fig.~\ref{picene_fig3}, 
which plots the change in the Fermi surface as the 
lattice constants are reduced down to 96\%.  
With the updated experimental lattice
constants the Fermi surface at ambient pressure 
is remarkably simpler.  We have however the sensitivity 
to the dopant positions: 
Namely, K$_{3}$Picene has a composite of a 
two-dimensional cylinder and a pair of one-dimensional sheets, 
while K$_2$K$_1$Picene has two pairs of more or less 
one-dimensional sheets.  
For both K3- and K2K1-structures, the pressure induces some
complexity in the Fermi surface structures. 
Remarkably, we can see that the topology of the Fermi 
surface changes, i.e., a Lifshitz transition 
occurs as the lattice is shrunk, for both of K3- and
K2K1-structures. It is an interesting future issue to investigate the
relation between Tc and the dopant arrangement, especially the 
Lifshitz transition. 

\begin{acknowledgements}
For the cuprates HA wishes to thank Hirofumi Sakakibara, 
Kazuhiko Kuroki, Ryotaro Arita and Douglas J. Scalapino 
for collaborations.  The study has been supported by a 
Grant-in-Aid for Scientific Research from JSPS
(No. 22340093).  For carbon systems 
we acknowledge Yuta Murakami, Philipp Werner and Naoto Tsuji 
for a collaboration (HA), and Erio Tosatti, Kosmas 
Prassides, Yoshihiro Kubozono and Takashi Kambe for discussions.  The 
study has been supported by 
LEMSUPER (JST-EU Superconductor Project) from JST.
\end{acknowledgements}


\begin{thebibliography}{}

\bibitem{aoki} 
Hideo Aoki, 
J. Superconductivity and Novel Magnetism {\bf 25}, 1243 (2012).

\bibitem{sakakibara} H. Sakakibara, H. Usui, K. Kuroki, R. Arita and H. Aoki, Phys. Rev. Lett. {\bf 105}, 057003 (2010); Phys. Rev. B {\bf 85}, 064501 (2012).

\bibitem{sakakibara_pressure} 
H. Sakakibara, K. Suzuki, H. Usui, 
K. Kuroki, R. Arita, D. J. Scalapino and H. Aoki, 
Phys. Rev. B {\bf 86}, 134520 (2012); H. Sakakibara et al, arXiv:1211.1805.

\bibitem{nishiguchi} K. Nishiguchi, K. Kuroki, R. Arita, 
T. Oka and H. Aoki, arXiv:1212.6320.


\bibitem{murakami} 
Y. Murakami, P. Werner, N. Tsuji and H. Aoki, arXiv:1305.5771.


 \bibitem{Mitsuhashi2010} R.~Mitsuhashi et al, 
Nature, {\bf 464}, 76 (2010); 
Y. Kubozono et al, 
Phys. Chem. Chem. Phys. {\bf 13}, 16476 (2011).

\bibitem{Jorgensen}
J.D. Jorgensen, D.G. Hinks, O.Chmaissem, D.N. Argyiou, J.F. Mitchell, and B. Dabrowski, in 
{\it Lecture Notes in Physics}, {\bf 475}, p.1 (1996).
\bibitem{Bianconi}
A. Bianconi, G. Bianconi, S. Caprara, D. Di Castro, 
H. Oyanagi and N. L. Saini, 
J. Phys.: Condens. Matter {\bf 12}, 10655 (2000); 
A. Bianconi, S. Agrestini, G. Bianconi, D. Di Castro, and 
N. L. Saini, J. Alloys Compd. {\bf 317-318}, 537 (2001); 
N. Poccia, A. Ricci and A. Bianconi, Adv. Condens. 
Matter Phys. {\bf 2010}, 261849 (2010). 

\bibitem{Maekawa}
S. Maekawa, J. Inoue and T. Tohyama, in {\it The Physics 
and Chemistry of Oxide Superconductors}, edited by Y. Iye 
and H. Yasuoka (Springer, Berlin, 1992), Vol. {\bf 60}, 
pp. 105-115. 
\bibitem{Andersen}
O.K. Andersen, A.I Liechtenstein, O. Jepsen, and F. Paulsen,
 J. Phys. Chem. Solids {\bf 56}, 1573 (1995).
\bibitem{Feiner}
L.F. Feiner, J.H. Jefferson and R. Raimondi, Phys. Rev. Lett. {\bf 76}, 4939 (1996).
\bibitem{Pavarini}
E. Pavarini, I. Dasgupta, T. Saha-Dasgupta, O. Jepsen, and O. K. Andersen, Phys. Rev. Lett. {\bf 87}, 047003 (2001).
\bibitem{Kotliar}
C. Weber, K. Haule, and G. Kotliar , 
Phys. Rev. B {\bf 82}, 125107(2010).

\bibitem{Weber}
C. Weber, C. -H. Yee, K. Haule and G. Kotliar, 
Eur. Phys. Lett. {\bf 100}, 37001 (2012). 
\bibitem{Takimoto}
T. Takimoto, T. Hotta and K. Ueda, Phys. Rev. B {\bf 69}, 104504 (2004).

\bibitem{Klehe}
A.-K. Klehe, A. K. Gangopadhyay, J. Diederichs and J. S. Schilling Physica {\bf 213C}, 266 (1993); {\bf 223C} 121(1994).
\bibitem{Gao}
L. Gao, Y. Y. Xue, F. Chen, Q. Xiong, R. L. Meng, D. Ramirez, C. W. Chu, J.H Eggert, and H.K. Mao, 
Phys. Rev. B {\bf 50}, 4260 (1994).

\bibitem{Hardy}
F. Hardy, N. J. Hillier, C. Meingast, D. Colson, Y. Li, N. Barisic, G. Yu, X. Zhao, M. Greven, and J. S. Schilling, Phys. Rev. Lett. {\bf 105}, 167002 (2010).
\bibitem{Gugenberger}
F. Gugenberger, C. Meingast, G.Roth, K. Grube, V. Breit, T. Weber, H. Wuhl, S. Uchida, and Y. Nakamura, Phys. Rev. B {\bf 49}, 13137 (1994).
\bibitem{Meingast}
C. Meingast, A. Junod and E. Walker, Physica C {\bf 272}, 106 (1996).


\bibitem{Schilling93} 
A. Schilling, M. Cantoni, J. D. Guo, and H. R. Ott, Nature {\bf 363},
56 (1993).


 \bibitem{c60_1}
Y. Takabayashi, A. Y. Ganin, P. Jeglic, D. Arcon, T. Takano, Y. Iwasa, Y. Ohishi, M. Takata, N. Takeshita, K. Prassides, and M. J. Rosseinsky, Science {\bf 323}, 1585 (2009).
  \bibitem{c60_2}
  M. Capone, M. Fabrizio, C. Castellani, and E. Tosatti, Rev. Mod. Phys. {\bf 81}, 943 (2009).
 \bibitem{c60_3}
O. Gunnarsson, Rev. Mod. Phys. {\bf 69}, 575 (1997).

\bibitem{1dimdmrg1}
M. Tezuka, R. Arita, and H. Aoki, Phys. Rev. Lett. {\bf 95}, 226401 (2005) ; Phys. Rev. B {\bf 76}, 155114 (2007).
\bibitem{1dimdmrg3}
H. Fehske, G. Hager and E. Jeckelmann, Europhys. Lett. {\bf 84}, 57001(2008).
\bibitem{sse1}
R. T. Clay and R. P. Hardikar, Phys. Rev. Lett. {\bf 95}, 096401(2005).


\bibitem{HHdmft1}
J. K. Freericks and M. Jarrell, Phys. Rev. Lett. {\bf 75}, 2570 (1995).
 \bibitem{HHdmft2}
J. Bauer, EPL {\bf 90}, 27002 (2010).
 \bibitem{HHdmft3}
J. Bauer and A. C. Hewson, Phys. Rev. B {\bf 81}, 235113 (2010).
\bibitem{HHdmft4}
P. Werner and A. J. Millis, Phys. Rev. Lett. {\bf 99}, 146404 (2007).
\bibitem{HHnrg}
W. Koller, D. Meyer, Y. Ono and A. C. Hewson, Europhys. Lett. {\bf66}, 559 (2004).
\bibitem{HHdmft5}
J. K. Freericks and M. Jarrell, Phys. Rev. Lett. {\bf 75}, 2570 (1995).
 \bibitem{HHdmft6}
G. Sangiovanni, M. Capone, C. Castellani, and M. Grilli, Phys. Rev. Lett. {\bf 94}, 026401 (2005).
 \bibitem{HHdmft7}
J. Bauer, J. E. Han, and O. Gunnarsson,  Phys. Rev. B {\bf 87}, 054507 (2013).

\bibitem{nowadnick} 
E. A. Nowadnick et al, Phys. Rev. Lett. {\bf 109}, 246404 (2012).

\bibitem{sc3}
  A. B. Migdal, Sov. Phys. JETP {\bf 7},  996 (1958).
\bibitem{sc4}
  G. M. Eliashberg,  Sov. Phys. JETP {\bf 11}, 696 (1960) ;  {\bf 12}, 1000 (1961).
\bibitem{sc5}
  W. L. McMillan, Phys. Rev. {\bf 125} 331 (1968).

 \bibitem{Kambe2012} T.~Kambe, X.~He, Y.~Takahashi, Y.~Yamanari,
	 K.~Teranishi, H.~Mitamura, S.~Shibasaki, K.~Tomita, R.~Eguchi,
	 H.~Goto, Y.~Takabayashi, T.~Kato, A.~Fujiwara, T.~Kariyado,
	 H.~Aoki, and Y.~Kubozono, Phys. Rev. B, {\bf 86}, 214507
	 (2012).
 \bibitem{Kosugi2011} T.~Kosugi, T.~Miyake, S.~Ishibashi, R.~Arita, and H.~Aoki,
	 Phys. Rev. B, {\bf 84}, 214506 (2011).
 \bibitem{QuantumEspresso} P.~Giannozzi et al, 
J.~Phys.:~Condens.~Matter, {\bf 21}, 395502 (2009).

\end{thebibliography}
\end{document}